\documentclass[12pt]{article}

\title{ Schr\"odinger group and quantum finance}

\begin{document}

\author{  Juan M. Romero\thanks{ jromero@correo.cua.uam.mx},  Ulises Lavana
\\
\it Departamento de Matem\'aticas Aplicadas y Sistemas,\\
\it Universidad Aut\'onoma Metropolitana-Cuajimalpa\\
\it M\'exico, D.F  01120, M\'exico \\
[0.6cm]
 Elio Mart\'inez Miranda\\
 \it  Facultad de Econom\'ia,\\
\it Universidad Nacional Aut\'onoma de M\'exico \\
\it M\'exico, D.F  04510, M\'exico }

\pagestyle{plain}
\date{}

\maketitle

\begin {abstract}

 Using the one dimensional free particle symmetries,  the quantum finance symmetries are obtained. Namely,    it is shown that Black-Scholes equation is invariant under Schr\"o\-dinger group. In order to do this,  the one dimensional free non-relativistic particle and its symmetries are revisited. To get  the Black-Scholes equation symmetries,  the particle mass is identified as  the inverse of square of the vola\-ti\-lity.  Furthermore,  using financial variables, a Schr\"odinger algebra representation is constructed.\\

\end{abstract}

\section{Introduction}

Lately,  mathematical techniques developed  in physics  have been employed to study systems from other areas. For example, 
the Black-Scholes \cite{black:gnus} and Merton \cite{merton:gnus} equation 
 is very important to study finance theory and it  can be mapped to the  Schr\"odinger equation   \cite{baaquie:gnus}.  
 Then  technics that arise  in quantum mechanics can be used to study financial phenomena, this fact allowed the birth of a new discipline, the so call Quantum Finance  \cite{baaquie:gnus}.  It is well known that symmetry groups are useful to study different systems. Given that the conformal  group is the   largest  symmetry group of special relativity \cite{maldacena2:gnus}, this group is very important  in  physics. Furthermore,  the  Schr\"odinger   group is a non-relativistic  conformal group  and     it is  the symmetry group for the free Schr\"odinger equation  \cite{hagen:gnus,hagen1:gnus}. It is worth  mentioning  that    in  1882  Sophus Lie  showed that the Fick equation, which describes diffusion,  is invariant under   the Schr\"odinger group  \cite{lie:gnus}.  \\
 
 In this paper, it will be shown that Black-Scholes equation is invariant under Schr\"odinger group. In order to do this,  the one dimensional free non-relativistic particle and its symmetries will be revisited. The quantum version of the free non-relativistic particle  and its symmetries will be revisited too. To get  the Black-Scholes equation symmetries,  the particle mass is identified as  the inverse of square of the volatility.  Furthermore,  using financial variables, a Schr\"odinger algebra representation is constructed.\\
 
This paper is organized as follows: in section $2$ a brief review about one dimensional non-relativistic free particle and its symmetries  is given; in section $3$ the one dimensional free Schr\"odinger equation and its symmetries are  studied; in section $4$ the Black-Scholes equation and its symmetries are studied.  Finally, in section $5$ a summary is given.

\section{Free particle }

The one dimensional non-relativistic free particle action is given by 
\begin{eqnarray}
S=\int dt \frac{m}{2}\left(\frac{dx}{dt}\right)^{2},\label{eq:action}
\end{eqnarray}
this is the simplest mechanics system. Now, if $\alpha, \beta, \gamma,\delta, a, v, c$ are constants,  the conformal  coordinate transformations
\begin{eqnarray}
t^{\prime}=\frac{\alpha t+\beta}{\gamma t+\delta},\qquad x^{\prime}=\frac{ax+vt+c}{\gamma t+\delta}, \qquad a^{2}=\alpha \delta -\beta \gamma\not = 0, 
\label{eq:conformes}
\end{eqnarray}
can be constructed. This coordinates transformation includes 
temporal translations 
\begin{eqnarray}
t^{\prime}=t+\beta, \qquad x^{\prime}=x,
\label{eq:tt1}
\end{eqnarray}
spatial translations 
\begin{eqnarray}
t^{\prime}, \qquad x^{\prime}=x+c,
\label{eq:tt2}
\end{eqnarray}
Galileo`s transformations  
\begin{eqnarray}
t^{\prime}, \qquad x^{\prime}=x+vt,\label{eq:tt3}
\end{eqnarray}

anisotropic scaling 
\begin{eqnarray}
 t^{\prime}=a^{2} t, \qquad x^{\prime}=ax
 \label{eq:tt4}
\end{eqnarray} 
and  the special conformal transformations
\begin{eqnarray}
t^{\prime}=\frac{1}{\gamma t+1},\qquad x^{\prime}=\frac{x}{\gamma t+1}.\label{eq:tt5}
\end{eqnarray}
Now, with  the  coordinates transformation (\ref{eq:conformes}), the action
\begin{eqnarray}
S^{\prime}=\int dt^{\prime} \frac{m}{2}\left(\frac{dx^{\prime}}{dt^{\prime}}\right)^{2}
\end{eqnarray}
can be defined, which  satisfies 
\begin{eqnarray}
S^{\prime}=S+\frac{m}{2}
\int dt \left(\frac{d \phi(x,t)}{dt}\right),
\end{eqnarray}
where 
\begin{eqnarray}
 \phi(x,t)= \frac{1}{a^{2}}\left(   2av x+ v^{2}t-\frac{\gamma \left(  ax +vt +c \right)^{2}}{ \gamma t+\delta }  \right).\label{eq:phace}
\end{eqnarray}
Then, the one dimensional non-relativistic free particle action  (\ref{eq:action}) is invariant under the coordinates transformation (\ref{eq:conformes}). \\
 
 It is  shown below that the conformal coordinate transformations (\ref{eq:conformes}) and the quantity (\ref{eq:phace}) are used to study Black-Schole equation symmetries, which is important in financial theory.

\subsection{ Conservative quantities}

For the  one dimensional non-relativistic particle, the  following  quantities 
\begin{eqnarray}
P&=&m\dot x,\label{eq:qfp1}\\
H&=&\frac{P^{2}}{2m},\\
G&=&tP-mx,\\
K_{1}&=& tH-\frac{1}{2} xP,\\
K_{2}&=& t^{2}H-t xP+\frac{m}{2}x^{2}\label{eq:qfp2}
\end{eqnarray}
are conserved.\\

The momentum  $P$ is associated with spatial  translations (\ref{eq:tt2}). The Hamiltonian $H$
is associated with temporal translations  (\ref{eq:tt1}) . The quantity $G$ is associated with Galileo`s transformations  (\ref{eq:tt3}).  
While $K_{1}$ is associated with anisotropic scaling  (\ref{eq:tt4})and $K_{2}$ is associated with the special conformal transformations (\ref{eq:tt5}).\\

Furthermore, using the Poisson brackets, it can be shown that the  following relations  
\begin{eqnarray}
\{P,H\}&=&0,\label{eq:poisson1}\\
\{P,K_{1}\}&=&\frac{1}{2}P,\\
\{P,K_{2}\}&=&G,\\
\{P,G\}&=&m,\\
\{H,K_{1}\}&=&H,\\
\{H,G\}&=&P,\\
\{H,K_{2}\}&=&2K_{1},\\
\{K_{1},K_{2}\}&=&K_{2},\\
\{K_{1},G\}&=&\frac{1}{2}G,\\
\{K_{2},G\}&=&0\label{eq:poisson2}
\end{eqnarray}
are satisfied.\\

\subsection{ Schr\"odinger group}

The  Schr\"odinger equation for the one dimensional non-relativistic free particle is  
\begin{eqnarray}
 i\hbar \frac{\partial\psi\left(x,t\right)}{\partial t} =
 -\frac{\hbar^{2}}{2m} \frac{\partial ^{2} \psi (x,t) }{\partial x^{2}}. \label{eq:schrodinger}
\end{eqnarray}
Now, if the particle is observed in a system with coordinates $\left(x^{\prime}, t^{\prime}\right),$ the particle has to be described by wave function 
 $\psi^{\prime} (x^{\prime},t^{\prime}),$ which satisfies 
\begin{eqnarray}
 i\hbar \frac{\partial\psi^{\prime}\left(x^{\prime},t^{\prime}\right)}{\partial t^{\prime}} =
 -\frac{\hbar^{2}}{2m} \frac{\partial ^{2} \psi^{\prime} (x^{\prime},t^{\prime}) }{\partial x^{\prime 2}}.
\end{eqnarray}

With a long but  straightforward calculation, it can be proved that the  Schr\"odinger equation  (\ref{eq:schrodinger}) is invariant under conformal coordinate transformations   (\ref{eq:conformes}), where   the wave function transforms as
\begin{eqnarray}
\psi^{\prime}\left(x^{\prime},t^{\prime}\right)=\left( \sqrt{ \gamma t +\delta }\right) e^{\frac{im}{2\hbar}\phi(x,t) }\psi(x,t), \label{eq:onda-conforme}
\end{eqnarray}
here $\phi(x,t)$ is given by  (\ref{eq:phace}).\\

The conformal  symmetry for free Schr\"odinger equation was found by 
Niederer and  Hagen in 1972 \cite{hagen1:gnus,hagen:gnus}. However, this  symmetry was obtained by S. Lie in 1882 while he was studying the Fick equation.
\cite{lie:gnus}.  \\

Furthermore, according to  quantum mechanics, the quantities  (\ref{eq:qfp1})-(\ref{eq:qfp2}) are represented by  operators 
\begin{eqnarray}
\hat P&=&-i\hbar \frac{\partial }{\partial x},\label{eq:opsch1} \\
\hat H&=&\frac{\hat P^{2}}{2m},\label{eq:hamiltonian}\\
\hat G&=&t\hat P-mx,\\
\hat K_{1}&=& t\hat H-\frac{1}{4}\left( x\hat P+\hat P x\right),\\
\hat K_{2}&=& t^{2}\hat H-\frac{t}{2}\left(  x\hat P+\hat Px\right)+\frac{m}{2}x^{2}.\label{eq:opsch2}
\end{eqnarray}
Now, whether  $\left[A, B\right]=AB-BA,$ the following algebra 
\begin{eqnarray}
\left[\hat P,\hat H\right]&=&0,\label{eq:sch1} \\
\left[\hat P,\hat K_{1}\right]&=&\frac{i\hbar }{2} \hat P,\\
\left[\hat P,\hat K_{2}\right]&=&i \hbar \hat G,\\
\left[\hat P,\hat G\right]&=&i\hbar m,\\
\left[\hat H,\hat K_{1}\right]&=&i\hbar \hat H,\\
\left[\hat H,\hat G\right]&=&i\hbar \hat P,\\
\left[\hat H,\hat K_{2}\right]&=&2i\hbar \hat K_{1},\\
\left[\hat K_{1},\hat K_{2}\right]&=&i\hbar \hat K_{2},\\
\left[\hat K_{1},\hat G\right]&=&\frac{i\hbar }{2}\hat G,\\
\left[\hat K_{2},\hat G\right]&=&0\label{eq:sch2}
\end{eqnarray}
is satisfied, which is similar to the algebra  (\ref{eq:poisson1})-(\ref{eq:poisson2}). The algebra (\ref{eq:sch1})-(\ref{eq:sch2}) is the so called  Schr\"odinger algebra of Schr\"odinger group. Now, if $\hat O$ is an operator its  evolution is given by 
\begin{eqnarray}
\frac{d\hat O}{dt}=\frac{\partial \hat O}{\partial t}+\frac{i}{\hbar}\left[\hat H,\hat O\right].
\end{eqnarray}
Using this last equation and the algebra (\ref{eq:sch1})-(\ref{eq:sch2}), it is possible to show that  operators  (\ref{eq:opsch1})-(\ref{eq:opsch2}) are conserved.\\

In the next section, it  will be shown that Schr\"odinger symmetry arises  in Black-Scholes equation too.\\

\section{The Black-Scholes equation}

The Black-Scholes equation is  \cite{black:gnus,merton:gnus} 
\begin{eqnarray}
\frac{\partial C(s,t)}{\partial t}=-\frac{\sigma^{2}}{2}s^{2} \frac{\partial^{2} C(s,t)}{\partial s^{2}}-rs \frac{\partial C(s,t)}{\partial s}+ rC(s,t), \label{eq:bs}
\end{eqnarray}
where $C$ is the price of a derivative,  $s$ is  the price of the stock, $\sigma$ is the volatility and  $r$ is the annualized risk-free interest rate. This equation
is a remarkable result in finance theory.

Amazingly, the Black-Scholes equation (\ref{eq:bs})  is equivalent to Schr\"odinger equation \cite{baaquie:gnus}. In fact, using  the change of variable
\begin{eqnarray}
s=e^{x}\label{eq:bchange}
\end{eqnarray}
in equation (\ref{eq:bs}),  the following   result
\begin{eqnarray}
\frac{\partial C(x,t)}{\partial t}=-\frac{\sigma^{2}}{2} \frac{\partial^{2} C(x,t)}{\partial x^{2}}+\left(\frac{\sigma^{2}}{2}-r\right) \frac{\partial C(x,t)}{\partial x}+ rC(x,t).
\end{eqnarray}
is gotten.  Additionally,  if 
\begin{eqnarray}
C(x,t)=e^{\left[\frac{1}{\sigma^{2}}\left(\frac{\sigma^{2}}{2}-r\right)x+\frac{1}{2\sigma^{2}}\left(\frac{\sigma^{2}}{2}+r\right)^{2}t \right] }\psi(x,t)
\end{eqnarray}
the following equation
\begin{eqnarray}
\frac{\partial \psi(x,t)}{\partial t}=-\frac{\sigma^{2}}{2} \frac{\partial^{2} \psi(x,t)}{\partial x^{2}} \label{eq:ssb}
\end{eqnarray}
is obtained, wich  is like  Schr\"odinger equation (\ref{eq:schrodinger}). Then, since the Schr\"odinger equation (\ref{eq:schrodinger})  is invariant under conformal transformation (\ref{eq:conformes}), 
the  equation (\ref{eq:ssb}) is invariant under the same transformations.  In this case the function  $\psi(x,t)$ transforms as  
\begin{eqnarray}
\psi^{\prime}\left(x^{\prime},t^{\prime}\right)=\left( \sqrt{\gamma t+\delta}\right)
e^{\frac{1}{2\sigma^{2}}\phi(x,t) }\psi(x,t),
\end{eqnarray}
where  $\phi(x,t)$ is given by  (\ref{eq:phace}). Notice that the particle mass $m$ is   changed for $1/\sigma^{2}.$

\subsection{Schr\"odinger group and Black-Scholes equation}

Using the change of variable  (\ref{eq:bchange}), the coordinates  transformations
can be written (\ref{eq:conformes}) as 
\begin{eqnarray}
t^{\prime}=\frac{\alpha t+\beta}{\gamma t +\delta}, \qquad s^{\prime}= e^{\left( \frac{bt+c}{\gamma t +\delta }\right)} s^{\left(\frac{a}{\gamma t +\delta}\right)}.
\label{eq:bchanges}
\end{eqnarray}
Through  a long but straightforward calculation, it can be demonstrated that Black-Scholes equation (\ref{eq:bs}) is invariant under this last transformations, where  the price  $C(s,t)$ transforms as
\begin{eqnarray}
C^{\prime}\left(s^{\prime},t^{\prime}\right)&=&\left(\sqrt{\gamma t +\delta }\right)s^{\Phi_{1}(s,t)} e^{\Phi_{2}(s,t)} C(s,t),
\end{eqnarray}
here 
\begin{eqnarray}
\Phi_{1}(s,t)&=& \frac{-2a^{2}\gamma\left(\frac{\sigma^{2}}{2}-r\right)t +2a\left(b\delta -\gamma c\right) +2a^{2}(a-\delta)\left(\frac{\sigma^{2}}{2}-r\right)}{ 2a^{2}\sigma^{2}\left(\gamma t+\delta\right)}  \nonumber\\
& & -\frac{\gamma a^{2} \left(\ln s\right)}{ 2a^{2}\sigma^{2}\left(\gamma t+\delta\right)} \nonumber
\end{eqnarray}
and
\begin{eqnarray}
\Phi_{2}(s,t)&= &\left(\frac{  a^{2}\left(\frac{\sigma^{2}}{2}+r\right)^{2}(\alpha -\delta) +2a^{2}b\left(\frac{\sigma^{2}}{2}-r\right)+b\left(b\delta -2\gamma c\right)  }{ 2\sigma^{2}a^{2}(\gamma t+\delta )}\right) t\nonumber\\
& &+\frac{ a^{2} \beta \left(\frac{\sigma^{2}}{2}+r\right)^{2}+2a^{2} \left(\frac{\sigma^{2}}{2}-r\right)c-\gamma c^{2} -\gamma a^{2}\left(\frac{\sigma^{2}}{2}+r\right)^{2}t^{2} }{ 2\sigma^{2}a^{2}(\gamma t+\delta )} .\nonumber
\end{eqnarray}
Now,  the   Black-Scholes equation  (\ref{eq:bs}) can be written as 
\begin{eqnarray}
\frac{\partial C(s,t)}{\partial t}= \hat {\bf H} C(s,t).
\end{eqnarray}
where 
\begin{eqnarray}
\hat {\bf H}=-\frac{\sigma^{2}}{2}s^{2} \frac{\partial^{2} }{\partial s^{2}}-rs \frac{\partial }{\partial s}+ r.
\end{eqnarray}
Moreover,  using the operator 
\begin{eqnarray}
\hat \Pi &=&-i s\frac{\partial }{\partial s}+\frac{i}{\sigma^{2}}\left(\frac{\sigma^{2}}{2}-r\right),\label{eq:pi}
\end{eqnarray}
the operator  $\hat {\bf H}$ can be rewritten as
\begin{eqnarray}
\hat {\bf H}&=&\frac{\sigma ^{2}}{2}\hat \Pi^{2}+\frac{1}{2\sigma^{2}}\left(\frac{\sigma^{2}}{2}+r\right)^{2}.
\end{eqnarray}
Notice that operator $\bf \hat H$ is similar to Hamiltonian operator $\hat H$  (\ref{eq:hamiltonian}), where the particle mass $m$ is associated with $1/\sigma^{2}.$\\ 

Additionally, using   the operator  (\ref{eq:pi}) it is possible construct quantities related with the non-relativistic free particle conserved quantities (\ref{eq:opsch1})-(\ref{eq:opsch2}). In fact,   
 operators
\begin{eqnarray}
\hat \Pi &=&-i s\frac{\partial }{\partial s}+\frac{i}{\sigma^{2}}\left(\frac{\sigma^{2}}{2}-r\right),\label{eq:1}\\
\hat {\bf H}_{0}&=&\frac{\sigma ^{2}}{2}\hat \Pi^{2},\\
\hat {\bf G}&=&t\hat \Pi-\frac{1}{\sigma^{2}}\ln s,\\ 
\hat {\bf K}_{1}&=& t\hat {\bf H}_{0}-\frac{1}{4}\left( \ln s \hat \Pi+\hat\Pi \ln s \right),\\
\hat {\bf K}_{2}&=& t^{2} \hat {\bf H}_{0}  -\frac{t}{2}\left(  \ln s\hat \Pi+\hat \Pi \ln s \right)+\frac{1}{2\sigma^{2}}\left(\ln  s\right) ^{2}  \label{eq:2}
\end{eqnarray}
can be proposed, which are similar to quantities (\ref{eq:opsch1})-(\ref{eq:opsch2}). Furthermore, using the relation  
\begin{eqnarray}
[\ln s, \hat \Pi] &=&i, 
\end{eqnarray}
 the  algebra
\begin{eqnarray}
\left[\hat \Pi,\hat {\bf H}\right]&=&0, \label{eq:3}\\
\left[\hat {\Pi},\hat {\bf K}_{1}\right]&=&\frac{i }{2} \hat \Pi,\\
\left[\hat \Pi,\hat {\bf K}_{2}\right]&=&i  \hat {\bf G},\\
\left[\hat \Pi,\hat {\bf G}\right]&=&\frac{i}{ \sigma^{2}},\\
\left[\hat {\bf H},\hat {\bf K}_{1}\right]&=&i \hat {\bf H}_{0}\\
\left[\hat {\bf H},\hat {\bf G}\right]&=&i  \hat \Pi,\\
\left[\hat {\bf H},\hat {\bf K}_{2}\right]&=&2i \hat {\bf K}_{1},\\
\left[\hat {\bf K}_{1},\hat {\bf K}_{2}\right]&=&i \hat {\bf K}_{2},\\
\left[\hat {\bf K}_{1},\hat {\bf G}\right]&=&\frac{i }{2}\hat {\bf G},\\
\left[\hat {\bf K}_{2},\hat {\bf G}\right]&=&0.\label{eq:4}
\end{eqnarray}
is satisfied. Then the operators (\ref{eq:1})-(\ref{eq:2}) satisfy the Schr\"odinger algebra.\\

Another study about Black-Scholes symmetries can be seen in \cite{gazizov:gnus}.

\section{Summary}
It was shown  that Black-Scholes equation is invariant under Schr\"odinger group. In order to do this,  the one dimensional free non-relativistic particle and its symmetries were  revisited.   The quantum version of the free non-relativistic particle  and its symmetries were  revisited too. To get  the Black-Scholes equation symmetries,  the particle mass was identified as  the inverse of square of the volatility. Besides,  using financial variables, a Schr\"odinger algebra representation was  constructed. This result shows that physical techniques can be employed to study other disciplines.

\end{document}